\documentclass[]{elsarticle}
\makeatletter
\def\ps@pprintTitle{%
	\let\@oddhead\@empty
	\let\@evenhead\@empty
	\def\@oddfoot{}%
	\let\@evenfoot\@oddfoot}
\makeatother
\usepackage{fancyhdr}

\usepackage{epsfig}
\usepackage{amsfonts}
\usepackage{amssymb, graphics, amsmath}
\usepackage{dsfont}
\usepackage{color}
\usepackage{pdflscape}
\usepackage{pifont}
\usepackage{graphicx} 
\usepackage{bm}
\usepackage{slashed}

\allowdisplaybreaks

\def\slashchar#1{\setbox0=\hbox{$#1$}           
	\dimen0=\wd0                                    
	\setbox1=\hbox{/} \dimen1=\wd1                  
	\ifdim\dimen0>\dimen1                           
	\rlap{\hbox to \dimen0{\hfil/\hfil}}            
	#1                                             
	\else                                          
	\rlap{\hbox to \dimen1{\hfil$#1$\hfil}}        
	/                                           
	\fi}       

\begin{document}
\title{Contraction Diagram Analysis in Pion-Kaon Scattering}

\author[bonn]{Chaitra Kalmahalli Guruswamy }
\author[bonn,fzj,tbilisi]{Ulf-G. Mei\ss{}ner}
\author[bonn]{Chien-Yeah Seng}

\address[bonn]{Helmholtz-Institut f\"ur Strahlen- und
	Kernphysik and Bethe Center for Theoretical Physics,\\
	Universit\"at Bonn, D-53115 Bonn, Germany}
\address[fzj]{Institute for Advanced Simulation, Institut f\"ur Kernphysik and
	J\"ulich Center for Hadron Physics,\\ Forschungszentrum J\"ulich,
	D-52425 J\"ulich, Germany}
\address[tbilisi]{Tbilisi State University, 0186 Tbilisi, Georgia}

\begin{abstract}
We study the contributions from the connected and disconnected contraction diagrams to the pion-kaon scattering amplitude
within the framework of SU$(4|1)$ partially-quenched chiral perturbation theory. Combining this with a finite-volume analysis,
we demonstrate that a lattice calculation of the easier computable connected correlation functions is able to provide valuable
information of the noisier disconnected correlation functions, and may serve as a theory guidance for the future refinement
of the corresponding lattice techniques.
\end{abstract}

\maketitle

\thispagestyle{fancy}

\section{Introduction}

Pion-kaon ($\pi K$) scattering is the simplest hadronic scattering process that involves a strange quark,
and therefore it plays a crucial role in our understanding of the SU(3) chiral symmetry breaking of the
Quantum Chromodynamics (QCD)~\cite{Gasser:1984gg}.
The $\pi K$ scattering amplitude was calculated within the framework of Chiral Perturbation Theory (ChPT)
at one loop \cite{Bernard:1990kw,Bernard:1990kx} and at two loops \cite{Bijnens:2004bu},
with the appearance of certain low-energy constants (LECs), some of which can be fixed in other processes. Naturally,
this also provides
motivation for the study of $\pi K$ scattering using one of the standard first-principle treatments of the strong
interaction, namely  lattice QCD\footnote{For investigations of $\pi K$ scattering using dispersion relations,
  see e.g.~\cite{Buettiker:2003pp,Pelaez:2016klv}.}.

Furthermore, due to the similar isospin structures, an improved understanding of 
$\pi K$ scattering also provides useful insights for  $\pi N$ scattering, which is an important ingredient towards
resolving the current disagreement between the
lattice~\cite{Bali:2016lvx,Yang:2015uis,Abdel-Rehim:2016won,Yamanaka:2018uud,Durr:2015dna} and the
dispersion-theoretical~\cite{Hoferichter:2015dsa,Hoferichter:2015tha,Hoferichter:2016ocj,RuizdeElvira:2017stg}
determinations of the pion-nucleon sigma term.

So far there exists a number of exploratory studies of $\pi K$ scattering, in both the $I=3/2$ and $I=1/2$ channels
\cite{Miao:2004gy,Beane:2006gj,Nagata:2008wk,Fu:2011wc,Lang:2012sv,Sasaki:2013vxa,Wilson:2014cna,Helmes:2018nug}. The $I=1/2$ channel is
of much interest as it provides useful information about the $K^*$ resonance, but it turns out that this channel is
much more difficult to handle on the lattice, due to the existence of correlation functions involving the contraction of
one or more pairs of quarks at the same temporal point (which are often called ``disconnected diagrams''). Such diagrams
have low signal-to-noise ratio, and are also the main reason for the increased difficulty in the lattice study of
$\pi\pi$ scattering at lower isospin. Obviously, one cannot claim to have a controlled error analysis in the lattice study
of $\pi K$ scattering without properly understanding the contribution from the disconnected diagrams.

The recent years have seen a systematic development of a theory analysis of contraction diagrams in hadron-hadron
interactions based on Partially-Quenched Chiral Perturbation Theory (PQChPT). The underlying principle is rather
straightforward: Contraction diagrams that are inseparable in a physical amplitude would become separable upon the
introduction of extra quark flavors. Since this separation is unphysical, it will unavoidably involve new parameters
that cannot be fixed by experiment, but can be determined from lattice simulations. This method was successfully
applied in the analysis of $\pi\pi$ scattering~\cite{Guo:2013nja,Acharya:2017zje,Acharya:2019meo} and the parity-odd
$\pi N$ coupling~\cite{Guo:2018aiq}. In this paper we generalize it to $\pi K$ scattering in both the finite and infinite
volume. We demonstrate that, from the lattice calculation of the two easier computable connected diagrams in the $I=3/2$ channel,
one acquires enough information to make definite predictions of the exponential behavior of the harder to compute, disconnected
diagram in $I=1/2$. This provides a useful theory gauge to the calculation of the latter on lattice. For a discussion
of the status of various scattering processes pertinent to chiral dynamics in the continuum and on the lattice,
see~\cite{Meissner:2019bsu}.

This work is organized as follows. In Sec.~2 we introduce the different contraction diagrams in $\pi K$ scattering,
and demonstrate how they can be expressed in terms of physical scattering amplitudes in a deformation of QCD with an
extended flavor sector. In Sec.~3 we introduce SU$(4|1)$ PQChPT in the infinite volume, and use it to calculate the different
contraction diagrams up to one-loop accuracy, $O(p^4)$. In Sec.~4 we discuss the implications of the results above to the actual
lattice calculations which are carried out in a finite volume. The final conclusions are given in Sec.~5.

\section{Contraction diagrams in $\pi K$ scattering}
Assuming isospin symmetry, the $\pi K$ scattering amplitude can be categorized into two isospin channels,
$T^{3/2}$ and $T^{1/2}$. In particular, $T^{3/2}(s,t,u)$ is given by the  scattering amplitude of
$\pi^{+}(k_{1}) K^{+}({p_1})\to\pi^{+}(k_{2}) K^{+}({p_2})$, where the Mandelstam variables $s,t,u$ are defined
as $s=(k_{1}+p_1)^{2}$, $t=(k_1-k_{2})^2$, $u=(k_1-p_2)^{2}$, respectively, subject to the constraint $s+t+u=2(M_\pi^2+M_K^2)$.
The $I=1/2$ amplitude can be obtained from $T^{3/2}$ by appropriate crossing:
\begin{figure}
	\begin{center}
	\includegraphics{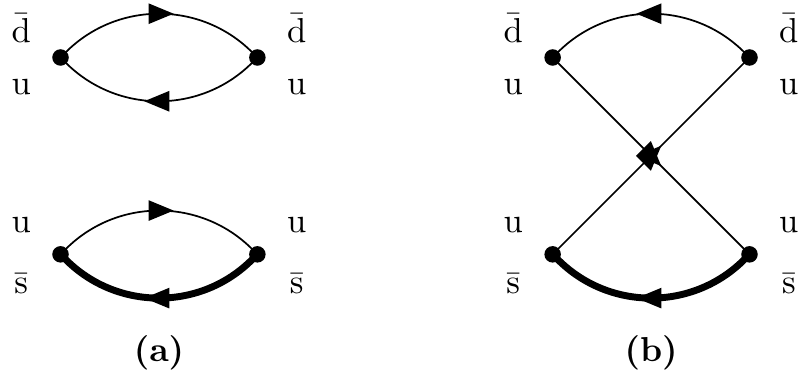}
		\caption{The quark contraction diagrams for $I=3/2$ $\pi K$ scattering. The amplitude for diagram \textbf{(a)}
			and \textbf{(b)} is given by $T_{a}(s,t,u)$ and $T_{b}(s,t,u)$ respectively. The thick line indicates
			the $\left\langle s\bar{s}\right\rangle$ contraction. The time flows in the horizontal direction.
			\label{fig:3/2contract}}
	\end{center}
\end{figure}
\begin{equation}
T^{1/2}(s,t,u)  =  \frac{3}{2}T^{3/2}(u,t,s) - \frac{1}{2}T^{3/2}(s,t,u)~.
\end{equation}

\begin{figure}
	\centering
	\includegraphics{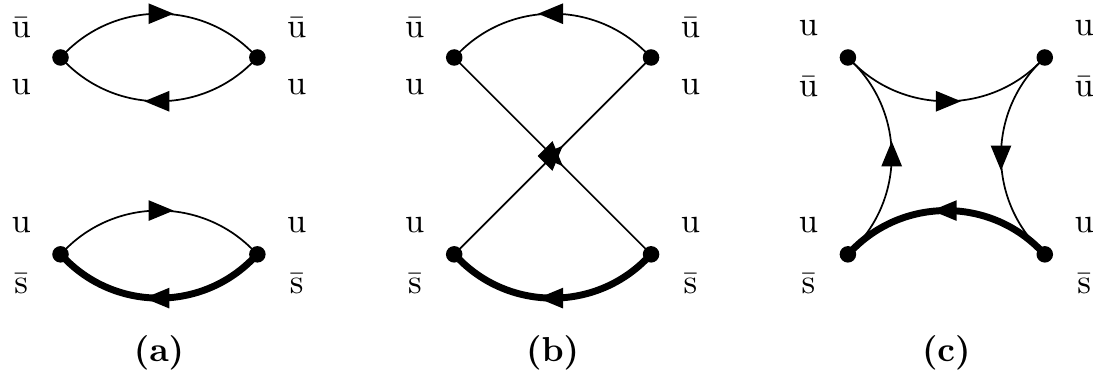}
	\caption{The contraction diagrams in the $I=1/2$ $\pi K$ scattering.\label{fig:1/2contract}}
\end{figure}

To construct interpolators of mesons on the lattice, one expresses the meson fields in terms of their ``constituent quarks'',
for example, $\pi^{+} = u\bar{d}$ and $K^{+} = u\bar{s}$. A lattice study of meson-meson scattering then consists
of computing correlation functions involving all possible contractions between quark and anti-quark pairs.
For instance, the $I=3/2$ amplitude represents the sum of the two independent contraction diagrams 
$T_{a}(s,t,u)$ and $T_{b}(s,t,u)$ depicted in Fig.~\ref{fig:3/2contract}: 
\begin{equation}
T^{3/2}(s,t,u) = T_{a}(s,t,u) + T_{b}(s,t,u).
\end{equation}
Both contraction diagrams above are purely connected, as there is no contraction between the quark--anti-quark pair at
the same time coordinate. Therefore, they are rather straightforwardly calculable on the lattice. The situation for
the $I=1/2$ amplitude is quite different. It involves three types of contraction diagrams displayed in
Fig.~\ref{fig:1/2contract}:
\begin{equation}
T^{1/2}(s,t,u)=T_{a}(s,t,u)-\frac{1}{2}T_b(s,t,u)+\frac{3}{2}T_c(s,t,u)~,
\end{equation}
among which the diagram~(c) contains a pair of disconnected contractions and is much noisier on the lattice. However,
from the theory point of view, $T_c$ is nothing but the $s\to u$ crossing of $T_b$ and is no more complicated than the latter.
Therefore, a precise theory description of the individual connected diagrams will automatically provide useful
information of the disconnected ones which can be directly contrasted to lattice results. 

In an ordinary three-flavor QCD the two connected diagrams in Fig.~\ref{fig:3/2contract} are inseparable in any
physical scattering amplitude, so one cannot study $T_b(s,t,u)$ by itself. The separation is possible, however, in a deformation
of QCD with an extended quark sector. In a generic meson-meson scattering, in order to isolate each contraction diagram
one requires a minimum number of four fermionic quarks~\cite{Guo:2013nja}.
But at the same time one needs also one ``bosonic quark", such that
its loop effect cancels with that from the extra fermionic quark, and thus to keep the sea dynamics identical to that
of ordinary three-flavor QCD. This leads to SU$(4|1)$ Partially-Quenched QCD (PQQCD), in which the quark
sector reads $q=(u,d,s,j;\tilde{j})$, where the first four quarks are fermionic and the last is bosonic. The quark mass
matrix is given by ${\cal M}=\mathrm{diag}(\bar{m},\bar{m},m_{s},\bar{m};\bar{m})$, where $m_s$ is the strange quark mass and
$\bar{m}<m_s$. Notice that this extended theory is actually simpler than that needed in the analysis of $\pi\pi$
scattering~\cite{Acharya:2017zje,Acharya:2019meo}. There, one needs again four fermionic quarks for the diagram separations,
but two bosonic quarks in order to keep the sea dynamics identical to a two-flavor QCD. That leads to an SU$(4|2)$ PQQCD
which has more pseudo-Nambu-Goldstone (pNG) particles than SU$(4|1)$ (see discussions in the next section).
The two contractions $T_a$ and $T_b$ can now be expressed in terms of physical scattering amplitudes in the extended theory:
 \begin{eqnarray}
T_{a}(s,t,u) & = & T_{(u\bar{s})(d\bar{j})\rightarrow(u\bar{s})(d\bar{j})}(s,t,u)~, \nonumber \\
T_{b}(s,t,u) & = & T_{(u\bar{s})(d\bar{j})\rightarrow(d\bar{s})(u\bar{j})}(s,t,u)~.\label{eq:TaTb}
\end{eqnarray}
 
\section{Analysis in SU$(4|1)$ PQChPT}

The right-hand side of Eq.~\eqref{eq:TaTb} can be calculated in the low-energy effective field theory (EFT) of
SU$(4|1)$ PQQCD, namely the SU$(4|1)$
PQChPT~\cite{Bernard:1993sv,Sharpe:1999kj,Sharpe:2000bc,Sharpe:2001fh,Bernard:2013kwa,Sharpe:2006pu,Golterman:2009kw}.
In this section we summarize the most important results relevant to this work, while interested readers may refer to
the literature cited above for more details.

Firstly, in complete analogy to the ordinary ChPT, the spontaneous chiral symmetry breaking $\mathrm{SU}(4|1)_{L}
\otimes \mathrm{SU}(4|1)_{R}\rightarrow \mathrm{SU}(4|1)_{V}$ in SU$(4|1)$ PQQCD generates pNG particles that are
expressed collectively in the following matrix-valued field:
\setlength{\abovedisplayskip}{10pt}
\begin{equation}
\mathbf{\Phi} = \begin{pmatrix}
\phi& \eta_{1} \\ 
\eta_{2}& \tilde{\phi}
\end{pmatrix},
\end{equation} 
with:
\begin{equation}
\phi =\begin{pmatrix}
u\bar{u} & u\bar{d} & u\bar{s} & u\bar{j} \\ 
d\bar{u} & d\bar{d} & d\bar{s} & d\bar{j} \\ 
s\bar{u} & s\bar{d} & s\bar{s} & s\bar{j} \\ 
j\bar{u}& j\bar{d} & j\bar{s} & j\bar{j}
\end{pmatrix}~,~~\eta_{1} =\begin{pmatrix}
u\bar{\tilde{j}}	\\ 
d\bar{\tilde{j}}	\\ 
s\bar{\tilde{j}}	\\ 
j\bar{\tilde{j}}	
\end{pmatrix}~,~~\eta_{2}= \begin{pmatrix}
\tilde{j}\bar{u}	&\tilde{j}\bar{d}  & \tilde{j}\bar{s} & \tilde{j}\bar{j}
\end{pmatrix}~,~~\tilde{\phi} = \tilde{j}\bar{\tilde{j}}~.
\end{equation}
The supertrace $(\mathrm{Str})$ of $\mathbf{\Phi}$ is defined as,
\begin{equation}
\mathrm{Str}\mathbf{\Phi} = \sum_{i=1}^{4}\phi_{ii}-\tilde{\phi}.
\end{equation}
Since we know that there are only $5^2-1=24$ independent pNG particles in $\mathbf{\Phi}$ (in contrast to the $6^2-1=35$
pNG particles in SU$(4|2)$ for $\pi\pi$ scattering), it is  more convenient to introduce a supertraceless matrix
$\mathbf{\Phi'}=\mathbf{\Phi}-\frac{1}{3} \mathrm{Str}\mathbf{\Phi}$. In particular, the diagonal components in
$\mathbf{\Phi'}$ give rise to four independent neutral pNG bosons $\{\pi_{0},\eta,\sigma_{a},\sigma_{b}\}$
by writing:
\begin{equation}
  (\mathbf{\Phi'})_{\mathrm{diag}}= \pi_{0}\lambda'_{3}+\sigma_{a}\lambda'_{8} + \frac{1}{2\sqrt{2}}(3\eta-\sigma_{b})\lambda'_{15}
  +  \frac{1}{2\sqrt{2}}(-\eta + 3\sigma_{b})\lambda'_{24}\label{eq:diagonal}
\end{equation}
where 
\begin{equation*}
\lambda'_{3} = \frac{1}{\sqrt{2}}\mathrm{diag}(1,-1,0,0;0) ,\:\:\:  \lambda'_{8} = \frac{1}{\sqrt{6}}\mathrm{diag}(1,1,0,-2;0),
\end{equation*}

\begin{equation}
\lambda'_{15} =  \frac{1}{\sqrt{12}}\mathrm{diag}(1,1,-3,1;0),\:\:\: \lambda'_{24} =  -\frac{1}{\sqrt{24}}\mathrm{diag}(1,1,1,1;4).
\end{equation}
With this we can define the standard non-linear representation of the pNG particles,
\begin{equation}
U = \exp\left(\frac{\sqrt{2}i  \mathbf{\Phi'}}{F_{0}}\right),
\end{equation}
where $F_0$ is the pNG boson decay constant in the chiral limit, and proceed to construct the most general
effective chiral Lagrangian. At O($p^{2}$) we get:
\begin{equation}
  \mathcal{L}^{(2)} =\dfrac{ F^{2}_{0}}{4} \mathrm{Str}[\partial_{\mu}U^{\dagger}\partial^{\mu}U]
  + \dfrac{F^{2}_{0}}{4} \mathrm{Str}[\chi U^{\dagger}+U \chi^{\dagger}],\label{eq:Op2}
\end{equation}
where $\chi=2B_0{\cal M}$, with ${\cal M}$ the quark mass matrix.
Expanding Eq.~\eqref{eq:Op2} up to the quadratic terms of pNG fields, we find that there are no
mixing terms between different fields, and thus all the 24 pNG fields are indeed independent particles, with the
leading order (LO) squared masses given by one of the three following mass parameters:
\begin{equation}
  \mathring{M}_\pi^2 =2B_0\bar{m},\:\:\:\mathring{M}_K^2 = B_0(\bar{m}+m_s),\:\:\:\mathring{M}_\eta^2=\frac{2}{3}B_0(\bar{m}+2m_s),
\end{equation}
satisfying the Gell-Mann-Okubo formula, $3\mathring{M}_\eta^2=4\mathring{M}_K^2-\mathring{M}_\pi^2$.
In particular, the four neutral particles $\{\pi_{0},\eta,\sigma_{a},\sigma_{b}\}$ have LO masses
$\{\mathring{M}_{\pi},\mathring{M}_{\eta},\mathring{M}_{\pi},\mathring{M}_{\pi}\}$, respectively.
One also finds that the pNG field propagators are given by the standard form:
\begin{equation}
S_\phi(k)=\frac{i}{k^2-M_\phi^2+i\epsilon},
\end{equation}
except that the $\sigma_b$ propagator acquires an extra negative sign.
In short, the diagonalization procedure of the neutral particles in Eq.\eqref{eq:diagonal} completely avoids
the cumbersome double-pole structures in the usual discussions of PQChPT propagators, and greatly
simplifies the one-loop analysis.

\begin{table}
	\caption{Coefficients of the UV divergence in the SU$(4|1)$ PQChPT.}
	\begin{center}
		\begin{tabular}{|c|c|c|c|c|c|c|c|c|c|}
			\hline 
			\rule[-1ex]{0pt}{2.5ex} i & 0 & 1 & 2 & 3 & 4 & 5 & 6 & 7 & 8 \\ 
			\hline 
			\rule[-1ex]{0pt}{2.5ex} $\Gamma_{i}$ & $\frac{1}{16}$ & $\frac{3}{32}$ & $\frac{3}{16}$  & 0 & $\frac{1}{8}$ & $\frac{3}{8}$ & $\frac{11}{144}$ & 0 & $\frac{5}{48}$ \\ 
			\hline 
		\end{tabular} \\
	\end{center}
\end{table}

Applying the Lagrangian in Eq.~\eqref{eq:Op2} at one loop results in ultraviolet (UV) divergences that are regulated
using dimensional regularization (DR) and reabsorbed into the LECs of the most general O($p^{4}$) chiral
Lagrangian without external sources~\cite{Sharpe:2006pu,Giusti:2008vb}:
\begin{eqnarray}
\mathcal{L}^{(4)}& = &L_{0} \mathrm{Str}[(\partial_{\mu} U^{\dagger})(\partial_{\nu} U)(\partial^{\mu} U^{\dagger})(\partial^{\nu}U)]\nonumber \\ 
& + &(L_{1}-\frac{1}{2}L_{0}) \mathrm{Str}[(\partial_{\mu}U^{\dagger})(\partial^{\mu}U)] \mathrm{Str}[(\partial_{\nu}U^{\dagger})(\partial^{\nu}U)]\nonumber  \\ 
& + &(L_{2}-L_{0})\mathrm{Str}[(\partial_{\mu}U^{\dagger})(\partial_{\nu}U)] \mathrm{Str}[(\partial^{\mu}U^{\dagger})(\partial^{\nu}U)]\nonumber \\ 
& + &(L_{3}+2L_{0}) \mathrm{Str}[(\partial_{\mu}U^{\dagger})(\partial^{\mu}U)(\partial_{\nu}U^{\dagger})(\partial^{\nu}U)]\nonumber \\ 
& + &L_{4}\mathrm{Str}[(\partial_{\mu}U^{\dagger})(\partial^{\mu}U)]\mathrm{Str}[U^{\dagger}\chi+\chi^{\dagger}U]\nonumber  \\
& + &L_{5}\mathrm{Str}[(\partial_{\mu}U^{\dagger})(\partial^{\mu}U)(U^{\dagger}\chi+\chi^{\dagger}U)]\nonumber  \\ 
& + & L_{6}(\mathrm{Str}[U^{\dagger}\chi+\chi^{\dagger}U])^2 + L_{7}(\mathrm{Str}[U^{\dagger}\chi-\chi^{\dagger}U])^2\nonumber  \\ 
& + & L_{8}\mathrm{Str}[\chi U^{\dagger} \chi U^{\dagger} + \chi^{\dagger} U \chi^{\dagger} U]~.
\label{eq:Op4}
\end{eqnarray}
Here it is useful to notice that the LECs
$\{L_{i}\}_{i=1}^{8}$ are identical to those in the ordinary SU(3) ChPT~\cite{Gasser:1984gg}, and the only
new LEC is $L_0$. This can be seen by observing that Eq.~\eqref{eq:Op4} is equivalent to the $O(p^4)$ chiral
Lagrangian of the ordinary SU(3) ChPT at tree level as long as the involved particles are the ordinary SU(3)
pNG bosons. The renormalized LECs are defined by $L_{i}^{r} = L_{i} -\lambda\Gamma_{i}$, where
\begin{equation}
\lambda = -\frac{1}{32 \pi^{2}}\left(\frac{2}{4-d} + \mathrm{log} (4\pi) -\gamma_E +1 \right),
\end{equation}
with $\gamma_E$ the Euler-Mascheroni constant, and $d$ is the number of space-time dimensions.
The divergence ($\beta$-function) coefficients $\{\Gamma_{i}\}$ are summarized in Tab.~1. 

Below we quote the analytical results up to $O(p^4)$ needed in this work. First, the physical pion, kaon masses and
the pion decay constant are just the same as in ordinary ChPT~\cite{Gasser:1984gg}: 
\begin{eqnarray}
M_{\pi}^{2} & = &\mathring{M}_{\pi}^{2} \left[1+ \mu_{\pi}-\frac{\mu_{\eta}}{3}+
\frac{16 M_{K}^{2}}{F_{\pi}^{2}}(2 L_{6}^{r}-L_{4}^{r})+ \frac{8M_{\pi}^{2}}{F_{\pi}^{2}}(2L_{6}^{r}+2L_{8}^{r}-L_{4}^{r}-L_{5}^{r})\right]~,
\nonumber \\
M_{K}^{2}& = &\mathring{M}_{K}^{2} \left[1+\frac{2\mu_{\eta}}{3}+ \frac{8 M_{\pi}^{2}}{F_{\pi}^{2}}(2 L_{6}^{r}-L_{4}^{r})
  + \frac{8M_{K}^{2}}{F_{\pi}^{2}}(4L_{6}^{r}+2L_{8}^{r}-L_{4}^{r}-L_{5}^{r})\right]~,\nonumber\\
F_{\pi}& = & F_{0}\left[1-2\mu_{\pi}-\mu_{K}+\frac{8M_{K}^{2}}{F_{\pi}^{2}}L_{4}^{r}+\frac{4M_{\pi}^{2}}{F_{\pi}^{2}}(L_{4}^{r}
  +L_{5}^{r})\right]~,
\end{eqnarray}
where $\mu_{P}=(M_{P}^{2}/32 \pi^{2} F_{\pi}^{2}) \ln(M_{P}^{2}/\mu^{2})$, with $\mu$ the scale of dimensional regularization.

The two contraction diagrams $T_a$ and $T_b$, expressed as SU$(4|1)$ physical scattering amplitudes in
Eq.~\eqref{eq:TaTb}, are given up to $O(p^4)$ as: 
\begin{eqnarray}
  T_{a}(s,t,u) & = & \dfrac{\mu_{\pi}}{8 F_{\pi}^{2} M_{\pi}^{2}(M_{\pi}^{2}-M_{K}^{2})}[16 M_{K}^{4}M_{\pi}^{2}
    + 2 M_{K}^{2}(14M_{\pi}^{4}-7M_{\pi}^{2}t +t^2)\nonumber \\
& &+ M_{\pi}^{2}(16M_{\pi}^{4} -15M_{\pi}^{2}t -6s^2-4su-6u^2)]+\dfrac{\mu_{K}}{4F_{\pi}^{2} M_{K}^{2}(M_{K}^{2}-M_{\pi}^{2})}\nonumber \\
 & &[8M_{K}^{6} +8M_{K}^{4}(2M_{\pi}^{2}-t)+M_{K}^{2}(8M_{\pi}^{4}-8M_{\pi}^{2}t-3s^2 -2su-3u^2 )  \nonumber \\
 & &+ M_{\pi}^{2}t^{2}]+\dfrac{M_{\pi}^{2}\mu_{\eta}}{72 F_{\pi}^{2} M_{\eta}^{2}(M_{K}^{2}-M_{\pi}^{2})}[-32M_{K}^{4} +M_{K}^{2}(18t-4M_{\pi}^{2})
\nonumber \\
& & +9M_{\pi}^{2}t]+ \dfrac{t(M_{\pi}^{2}+t)}{8F_{\pi}^{4}}\bar{J}_{\pi \pi}(t) -\frac{M_{\pi}^{2}(8M_{K}^{2}-9t)}{72F_{\pi}^{4}}
  \bar{J}_{\eta \eta}(t) +\dfrac{t^{2}}{8 F_{\pi}^{4}}\bar{J}_{KK}(t) \nonumber \\
  & & +\dfrac{(M_{K}^{2}+M_{\pi}^{2}-s)^2}{4F_{\pi}^{4}} \bar{J}_{\pi K}(s) + \dfrac{(M_{K}^{2}+M_{\pi}^{2}-u)^2}{4F_{\pi}^{4}}
  \bar{J}_{\pi K}(u) \nonumber \\
  & &  + \dfrac{M_{\pi}^{2}(4 M_{K}^{2}-3t)}{12F_{\pi}^{4}}\bar{J}_{\pi \eta}(t) + \frac{8}{F_{\pi}^{4}}[3 M_{K}^{4}+ M_{K}^{2}
    (4M_{\pi}^{2}-2t) \nonumber \\
& & + 3M_{\pi}^{4}-2M_{\pi}^{2}t-s^2-su-u^2] L_{0}^{r} + \frac{8}{F_{\pi}^{4}}(t-2M_{K}^{2})(t-2M_{\pi}^{2})L_{1}^{r}\nonumber \\
  & & +\frac{4}{F_{\pi}^{4}}[(s-M_{\pi}^{2}-M_{K}^{2})^2+(u-M_{\pi}^{2}-M_{K}^{2})^{2}]L_{2}^{r} +\frac{8}{F_{\pi}^{4}}
      [M_{K}^{2}(t-4M_{\pi}^{2})\nonumber \\
        & & +M_{\pi}^{2}t]L_{4}^{r} +\dfrac{32 M_{K}^{2}M_{\pi}^{2}}{F_{\pi}^{4}}L_{6}^{r} + \dfrac{4 M_{K}^{2}M_{\pi}^{2}
        -9t(M_{\pi}^{2}+t)}{576 \pi^{2}F_{\pi}^{4}}
\end{eqnarray}
and
\begin{eqnarray}
  T_{b}(s,t,u) & = &  \dfrac{M_{K}^{2}+M_{\pi}^{2}-s}{2F_{\pi}^{2}} + \frac{\mu_{\pi}}{24F_{\pi}^{2}M_{\pi}^{2}(M_{\pi}^{2}
    -M_{K}^{2})}[-2M_{K}^{4}(9M_{\pi}^{2}+2t) \nonumber \\
& & + M_{K}^{2}(6M_{\pi}^{4}+3M_{\pi}^{2}(5t+6u)+ 8t^2 +4tu)+ M_{\pi}^{2}(24M_{\pi}^{4}-2M_{\pi}^{2} \nonumber \\
& & (4t+9u)-8t^2-7tu-6u^2)]+ \dfrac{\mu_{K}}{12F_{\pi}^{2}M_{K}^{2}(M_{K}^{2}-M_{\pi}^{2})}[30M_{K}^{6}\nonumber \\
& & +2M_{K}^{4}(9M_{\pi}^{2}-10t-12u) + M_{K}^{2}(3M_{\pi}^{2}(3t-4u)-2t^2+2tu+6u^2) \nonumber \\
& & +M_{\pi}^{2}t(-M_{\pi}^{2}+2t+u)]+\dfrac{\mu_{\eta}}{8F_{\pi}^{2}(M_{K}^{2}-M_{\pi}^{2})}[-18M_{K}^{4}+M_{K}^{2}(15t\nonumber \\
& & + 18u -10M_{\pi}^{2})-6M_{\pi}^{2}(t-u)-3tu-6u^2]+\frac{1}{12F_{\pi}^{4}}[2t^2+tu+4M_{\pi}^{4} \nonumber \\
& & + M_{K}^{2}(4M_{\pi}^{2}-t)-2M_{\pi}^{2}(3t+2u)]\bar{J}_{\pi \pi}(t)\nonumber \\
& & +\frac{1}{24 F_{\pi}^4}[4 M_{K}^4 + M_{K}^2(4 M_{\pi}^2 -3 t - 4 u) + t(-M_{\pi}^2 + 2t + u)]\bar{J}_{KK}(t)\nonumber \\
  & &  + \frac{1}{16 F_{\pi}^{4}}[-4M_{K}^{4} + M_{K}^{2}(8M_{\pi}^{2}-2t) -4M_{\pi}^{4} -2M_{\pi}^{2}t +tu +2u^{2}]
  \bar{J}_{\pi K}(u)  \nonumber \\
  & & +\frac{t}{16 F_{\pi}^{4}}\frac{(M_{K}^{2}-M_{\pi}^{2})^2}{u}\bar{J}_{\pi K}(u) -\dfrac{M_{\pi} ^{2}(4M_{K}^{2}-3t)}{12F_{\pi}^{4}}
  \bar{J}_{\pi \eta}(t) +\frac{1}{144F_{\pi}^{4}}[44M_{K}^{4} \nonumber \\
& & + M_{K}^{2}(56M_{\pi}^{2}-42t -48u)-4M_{\pi}^{4}+6M_{\pi}^{2}(t-4u)+9tu +18u^2]\bar{J}_{K \eta}(u)\nonumber \\
  & & -\dfrac{(M_{K}^{2}-M_{\pi}^{2})^{2}(16M_{K}^{2}-8M_{\pi}^{2}-t)}{144F_{\pi}^{4}}\dfrac{\bar{J}_{K \eta}(u)}{u}
  + \dfrac{(M_{K}^{2}-M_{\pi}^{2})^{4}}{72F_{\pi}^{4}u^{2}}(9\bar{\bar{J}}_{\pi K}(u)+\bar{\bar{J}}_{K \eta}(u))\nonumber \\
  & & -\dfrac{8}{F_{\pi}^{4}}[-s^2-su-u^2 +3M_{K}^{4}+M_{K}^{2}(4M_{\pi}^{2}-2t)+ 3M_{\pi}^{4}-2M_{\pi}^{2}t]L_{0}^{r}
  + \dfrac{2}{F_{\pi}^{4}}[(t-2M_{K}^{2})\nonumber \\
& & (t-2M_{\pi}^{2})+(u-M_{\pi}^{2}-M_{K}^{2})^{2}]L_{3}^{r}- \dfrac{4M_{\pi}^{2}(M_{K}^{2}-M_{\pi}^{2}+s)}{F_{\pi}^{4}}L_{5}^{r}\nonumber \\
  & &+\dfrac{16M_{K}^{2}M_{\pi}^{2}}{F_{\pi}^{4}}L_{8}^{r} + \dfrac{1}{384\pi^{2}F_{\pi}^{4}}[11 M_{K}^{4}+M_{K}^{2}(10M_{\pi}^{2}
    -9t-7u)+3M_{\pi}^{4}\nonumber \\
& & +M_{\pi}^{2}(5t-3u)-5t^2+tu+u^2]~.
\end{eqnarray}
Here, the two-point functions $\bar{J}_{PQ}$ and $\bar{\bar{J}}_{PQ}$ are defined as~\cite{GomezNicola:2001as}:
\begin{eqnarray}
  \bar{J}_{PQ}(s)&=&\frac{1}{32\pi^2}\left[2+\left(\frac{\Delta}{s}-\frac{\Sigma}{\Delta}\right)\ln\frac{M_Q^2}{M_P^2}
    -\frac{\nu(s)}{s}\ln\frac{\left[s+\nu(s)\right]^2-\Delta^2}{\left[s-\nu(s)\right]^2-\Delta^2}\right]\nonumber\\
  \bar{\bar{J}}_{PQ}(s)&=&\bar{J}_{PQ}(s)-\frac{s}{32\pi^2}\left[\frac{\Sigma}{\Delta^2}+2\frac{M_P^2M_Q^2}{\Delta^{3}}
    \ln\frac{M_Q^2}{M_P^2}\right],
\end{eqnarray}
where 
\begin{equation}
\Delta=M_P^2-M_Q^2,\:\:\Sigma=M_P^2+M_Q^2,\:\:\nu(s)=\sqrt{[s-(M_P+M_Q)^2][s-(M_P-M_Q)^2]}.
\end{equation}
The third contraction diagram is simply given by $T_c(s,t,u)=T_b(u,t,s)$.
Finally, we are also interested in their values at the threshold,  $s_{0} = (M_{K}+M_{\pi})^{2}$, $
t_{0}=0$ and $u_{0}=(M_{K}-M_{\pi})^{2}$, which are given by:
\begin{eqnarray}
(T_{a})_{\mathrm{thr}} & = & \dfrac{M_{K}^{2}M_{\pi}^{2}}{F_{\pi}^{2}(M_{K}^{2}-M_{\pi}^{2})}\left(\frac{9}{2}\mu_{\pi}-4\mu_{K}-\dfrac{8M_{K}^{2}+M_{\pi}^{2}}{18M_{\eta}^2}\mu_{\eta}\right) + \dfrac{M_{K}^{2}M_{\pi}^{2}}{F_{\pi}^{4}}(-48L_{0}^{r}+32L_{1}^{r}\nonumber \\
& & +32L_{2}^{r}-32L_{4}^{r}+32L_{6}^{r}+\bar{J}_{\pi K}(s_{0})+\bar{J}_{\pi K}(u_{0}))+\dfrac{M_{K}^{2}M_{\pi}^{2}}{144 \pi^{2}F_{\pi}^{4}}\\
(T_{b})_{\mathrm{thr}} & = &  - \dfrac{M_{K} M_{\pi}}{F_{\pi}^{2}} + \dfrac{\mu_{\pi}}{4F_{\pi}^{2}(M_{K}^{2}-M_{\pi}^{2})}[M_{K}^{4}+2M_{K}^{3}M_{\pi}+5M_{K}^{2}M_{\pi}^{2}-10M_{K}M_{\pi}^{3}] \nonumber \\ 
&  & + \dfrac{\mu_{K}}{2F_{\pi}^{2}(M_{K}-M_{\pi})}[2M_{K}^{3} + 2M_{K}^{2}M_{\pi}+M_{K}M_{\pi}^{2}-M_{\pi}^{3}] \nonumber \\
&  & -\dfrac{\mu_{\eta}}{4F_{\pi}^{2}(M_{K}^{2}-M_{\pi}^{2})}[3M_{K}^{4}+6M_{K}^{3}M_{\pi}+11M_{K}^{2}M_{\pi}^{2}-6M_{K}M_{\pi}^{3}]\nonumber \\
& & -\frac{1}{8F_{\pi}^{4}}[(M_{K}-M_{\pi})^2(M_{K}^{2}+6M_{K}M_{\pi}+M_{\pi}^{2})]\bar{J}_{\pi K}(u_{0}) - \dfrac{1}{72 F_{\pi}^{4}}[M_{K}^{4} \nonumber \\
& & + 4M_{K}^{3}M_{\pi}-42M_{K}^{2}M_{\pi}^{2}+4M_{K}M_{\pi}^{3}+M_{\pi}^{4}]\bar{J}_{K \eta}(u_{0}) + \dfrac{(M_{K}+M_{\pi})^{4}}{72F_{\pi}^{4}}(9\bar{\bar{J}}_{\pi K}(u_{0})\nonumber \\
&  & +\bar{\bar{J}}_{K \eta}(u_{0})) + \dfrac{8M_{K}^{2}M_{\pi}^{2}}{F_{\pi}^{4}}\left(6L_{0}^{r}+2L_{3}^{r}-\frac{M_{K}+ M_{\pi}}{M_{K}}L_{5}^{r} +2L_{8}^{r}\right) \nonumber \\
& & +\dfrac{(M_{K}+M_{\pi})^{2}(5M_{K}^{2}+M_{\pi}^{2})}{384\pi^{2}F_{\pi}^{4}}
\end{eqnarray}
and $(T_{c})_{\mathrm{thr}} $ is obtained by replacing $M_{\pi} \rightarrow -M_{\pi}$ (which also means $u_{0} \to
s_{0}$) in $(T_{b})_{\mathrm{thr}}$.

\section{Finite-Volume Analysis}

We now discuss the implications of the results above, which are obtained in a field theory at infinite volume, to the discrete
energies calculated on the lattice in a finite volume. The analysis in this section is a straightforward generalization of
that in Ref.~\cite{Acharya:2019meo}.

To do so we construct three effective single-channel scattering amplitudes using $T_a$, $T_b$ and $T_c$. First, consider
the $2\times 2$ scattering matrix between the asymptotic states $\left|\psi_1\right\rangle=\left|u\bar{s}\right\rangle\left|
d\bar{j}\right\rangle$ and $\left|\psi_2\right\rangle=\left|d\bar{s}\right\rangle\left|u\bar{j}\right\rangle$. Diagonalizing
this matrix gives two single-channel scattering amplitudes:
\begin{equation}
T_{\alpha}(s,t,u)=T_a(s,t,u)+T_b(s,t,u),\:\:T_{\beta}(s,t,u)=T_a(s,t,u)-T_b(s,t,u)~.
\end{equation}
In particular, $T_\alpha(s,t,u)=T^{3/2}(s,t,u)$. The third single-channel amplitude is simply:
\begin{equation}
T_{\gamma}(s,t,u)=T^{1/2}(s,t,u)=T_{a}(s,t,u)-\frac{1}{2}T_b(s,t,u)+\frac{3}{2}T_c(s,t,u)~.
\end{equation}
For each single-channel amplitude one could perform the partial-wave expansion in the center-of-mass (CM) frame:
\begin{equation}
T(s,t,u)=\sum_{l=0}^\infty (2l+1)T_l(E)P_l(\cos\theta),
\end{equation}
where $E=\sqrt{s}$ is the CM energy, $\theta$ is the scattering angle and $\{P_l(x)\}$ are the Legendre polynomials. The $l=0$ (i.e. S-wave) partial-wave amplitude is parameterized as:
\begin{equation}
T_0(E)=\frac{8\pi E}{p\cot\delta_0(E)-ip},
\end{equation}
where $p$ is the CM momentum and $\delta_0(E)$ is the S-wave phase shift. At small $p$ one performs the effective
range expansion:
\begin{equation}
p\cot\delta_0(E)=-\frac{1}{a_0}+\frac{1}{2}r_0p^2+\ldots~,
\end{equation}
which defines the S-wave scattering length $a_0$ and effective range $r_0$.
The S-wave scattering lengths of the three single-channel amplitudes above are given by:
\begin{eqnarray}
a_0^{\alpha}&=&-\frac{1}{8\pi\sqrt{s_0}}\left[\left(T_a\right)_\mathrm{thr}+\left(T_b\right)_\mathrm{thr}\right]~,\nonumber\\
a_0^{\beta}&=&-\frac{1}{8\pi\sqrt{s_0}}\left[\left(T_a\right)_\mathrm{thr}-\left(T_b\right)_\mathrm{thr}\right]~,\nonumber\\
a_0^{\gamma}&=&-\frac{1}{8\pi\sqrt{s_0}}\left[\left(T_a\right)_\mathrm{thr}-\frac{1}{2}\left(T_b\right)_\mathrm{thr}
  +\frac{3}{2}\left(T_c\right)_\mathrm{thr}\right]~.
\end{eqnarray}
In particular, $a_0^{\beta}$ is the only one among the three that depends on the unphysical LEC $L_0^r$. The latter does not affect the pNG boson masses and decay constants at $\mathcal{O}(p^4)$, but does contribute to the scattering parameters. 

The discrete energies $E$ extracted from lattice correlation functions at finite volume can be obtained by solving
the single-channel L\"{u}scher's formula~\cite{Luscher:1986pf} (see also Ref.~\cite{Lang:2012sv,Doring:2011vk}
for more discussions):
\begin{equation}
p\cot\delta_0(E)=\frac{2\pi}{L}\pi^{-3/2}\mathcal{Z}_{00}(1;q^2),\:\:\:q=p\frac{L}{2\pi},
\end{equation}
where $L$ is the lattice size and $\mathcal{Z}_{00}$ is the L\"{u}scher zeta-function. This gives the
discrete ground-state energies of the three channels as known functions of the scattering lengths and lattice size
$E_0^i=f(a_0^i,L)$, $(i=\alpha,\beta,\gamma)$. Therefore, if we define $C_i(\tau)$ ($i=a,b,c$) as the lattice
correlation function corresponding to the contraction diagram of type $i$ at the Euclidean time $\tau$, then
the following combinations of correlation functions decay as a single exponential at large $\tau$:
\begin{eqnarray}
C_a(\tau)+C_b(\tau)&\sim&A_\alpha \exp\{-E_0^\alpha\tau\}\nonumber~,\\
C_a(\tau)-C_b(\tau)&\sim&A_\beta \exp\{-E_0^{\beta}\tau\}\nonumber~,\\
C_a(\tau)-\frac{1}{2}C_b(\tau)+\frac{3}{2}C_c(\tau)&\sim&A_\gamma \exp\{-E_0^{\gamma}\tau\}~.
\end{eqnarray} 

Hence, through a single lattice calculation of the difference between $C_{a}$ and $C_{b}$ which both appear in the
$I=3/2$ channel, one is able to obtain $E_0^{\beta}$ and thus fix the unknown LEC $L_0^r$. After doing so, all
the three discrete energies $\{E_0^{\alpha},E_0^{\beta},E_0^{\gamma}\}$ are fully predictable given any set of
lattice parameters. This is beneficial in multiple ways. For instance, we know that the most difficult disconnected
correlation function $C_c(\tau)$ depends on three exponents:
\begin{equation}
  C_c(\tau)\sim -\frac{1}{6}A_\alpha\exp\{-E_0^{\alpha}\tau\}-\frac{1}{2}A_\beta\exp\{-E_0^\beta\tau\}
  +\frac{2}{3}A_\gamma\exp\{-E_0^{\gamma}\tau\}~,
\end{equation}
and all the three exponents are known functions of $\left\{\bar{m},m_s,L\right\}$. This provides a useful theory
gauge of the accuracy for lattice studies of $C_c(\tau)$, which directly tests the lattice techniques in handling
disconnected diagrams. {\color{black} Furthermore, once $L_0^r$ is fixed from lattice data, the SU$(4|1)$ chiral Lagrangian (without external sources) will be completely known to NLO, so it could be applied to the lattice study of other interesting hadronic processes such as the $\pi\pi\rightarrow K\bar{K}$ scattering.}

We end this section by estimating the LEC $L_0^r$. Integrating out the strange quark from the theory, the graded
algebra SU$(4|1)$ reduces to SU$(3|1)$. The LEC $L_0$ appearing in this reduced theory should be the same as in the
SU$(4|2)$ version and thus is known with a sizeable uncertainty, $L_0^r = 1.1(1.0)\cdot10^{-3}$, {\color{black}which comes from an NNLO analysis of lattice data in Ref.~\cite{Boyle:2015exm}. In our new formalism, $L_0^r$ appears at NLO in the SU$(4|1)$ scattering amplitudes, so one could in general expect an order-of-magnitude improvement of its accuracy through the analysis of the lattice $\pi K$ contraction diagrams, just like what happened to the combination $3L_0^r+L_3^r$ in SU$(4|2)$ as demonstrated in Ref.~\cite{Acharya:2019meo}. The actual analysis will appear in a future work.}

\section{Conclusions}

As a natural generalization of previous works, we perform a PQChPT analysis of the different contraction diagrams
in $\pi K$ scattering, in both infinite and finite volume. We show that up to $\mathcal{O}(p^4)$ there is only one
undetermined LEC in the EFT, which can be fixed by the lattice study of the connected contraction diagrams in the $I=3/2$
channel. After doing so, the large-$\tau$ behavior of the disconnected correlation function in the $I=1/2$ channel
is predictable as a function of lattice parameters. These theory predictions can then be contrasted with actual
lattice calculations, serving as an important cross-check to the latter and reducing their associated systematic uncertainties.
Finally, the theory analysis above can be generalized  to $\pi N$ scattering and will be carried out in a follow-up work.

\vspace{0.5cm}

\section* {Acknowledgements}
We thank Feng-Kun Guo for some useful discussion.
This work is supported in part by  the DFG (Grant No. TRR110)
and the NSFC (Grant No. 11621131001) through the funds provided
to the Sino-German CRC 110 ``Symmetries and the Emergence of
Structure in QCD", and also by the Alexander von Humboldt Foundation through the Humboldt Research Fellowship (CYS). The work of UGM was also supported by the Chinese Academy of Sciences (CAS) through a President's
International Fellowship Initiative (PIFI) (Grant No. 2018DM0034) and by the VolkswagenStiftung (Grant No. 93562).

\end{document}